\documentclass[useAMS,usenatbib]{mn2e}
\usepackage[]{graphicx}
\usepackage[fleqn]{amsmath}
\usepackage{bm}
\usepackage{natbib}
\usepackage{textcomp}
\usepackage{url}
\usepackage{xcolor}
\usepackage{color}
\usepackage{soul}
\usepackage{tabularx}
\usepackage{xr}
\newcommand \mic{\mbox{$\mu${\rm m}}}

\definecolor{refcol}{rgb}{0.,0.,0.}

\title[GeMS review I]{Gemini multi-conjugate adaptive optics system review \nolinebreak I: \linebreak Design, trade-offs and integration}

\author[F.~Rigaut, B.~Neichel et al.]
{\parbox{\textwidth}{Fran\c{c}ois Rigaut,$^{1}$\thanks{E-mail: \texttt{francois.rigaut@anu.edu.au}}
Benoit Neichel,$^{2}$ 
Maxime Boccas,$^{2}$ 
C\'eline d'Orgeville,$^{1}$ 
Fabrice Vidal,$^{2}$ 
Marcos A. van Dam,$^{2,3}$
Gustavo Arriagada,$^{2}$ 
Vincent Fesquet,$^{2}$
Ramon L. Galvez,$^{2}$
Gaston Gausachs,$^{2}$ 
Chad Cavedoni,$^{2}$ 
Angelic W. Ebbers,$^{2}$
Stan Karewicz,$^{2}$
Eric James,$^{2}$ 
Javier L\"{u}hrs,$^{2}$
Vanessa Montes,$^{2}$
Gabriel Perez,$^{2}$ 
William N. Rambold,$^{2}$ 
Roberto Rojas,$^{2}$
Shane Walker,$^{2}$
Matthieu Bec,$^{4}$
Gelys Trancho,$^{4}$
Michael Sheehan,$^{4}$ 
Benjamin Irarrazaval,$^{4}$
Corinne Boyer,$^{5}$ 
Brent L. Ellerbroek,$^{5}$ 
\mbox{Ralf Flicker \textdied}, 
Damien Gratadour,$^{6}$
Aurea Garcia-Rissmann,$^{7}$ 
Felipe Daruich$^{8}$}\vspace{0.4cm}\\ 
\parbox{\textwidth}{$^{1}$the Australian National University, RSAA, Mount Stromlo Observatory, Cotter Road, Weston Creek ACT 2611, Australia\\
$^{2}$Gemini Observatory, c/o AURA, Casilla 603, La Serena, Chile\\
$^{3}$Flat Wavefronts, PO BOX 1060, Christchurch 8140, New Zealand\\
$^{4}$Giant Magellan Telescope Organization Corporation, PO box 90933, Pasadena, 
CA, 91109, USA\\
$^{5}$Thirty Meter Telescope, 1200 E. California Blvd., Pasadena, 
CA 91125, USA\\
$^{6}$LESIA, Observatoire de Paris-Meudon, place Jules Janssen, 
92190 MEUDON, France\\
$^{7}$European Southern Observatory, Karl-Schwarzschild-Str. 2, 
85748 Garching bei M\"unchen, Germany\\
$^{8}$ALMA, Alonso de C\'ordova 3107, Vitacura, Santiago, Chile}}

\date{Accepted 2013 October 22.  Received 2013 October 21; in original form 2013 September 27}

\pagerange{\pageref{firstpage}--\pageref{lastpage}} \pubyear{2013}

\def\LaTeX{L\kern-.36em\raise.3ex\hbox{a}\kern-.15em
    T\kern-.1667em\lower.7ex\hbox{E}\kern-.125emX}

\begin{document}

\label{firstpage}

\maketitle

\begin{abstract}
The Gemini Multi-conjugate adaptive optics System (GeMS) at the Gemini South telescope in Cerro Pach{\'o}n is the first sodium-based multi-Laser Guide Star (LGS) adaptive optics system. It uses five LGSs and two deformable mirrors to measure and compensate for atmospheric distortions. The GeMS project started in 1999, and saw first light in 2011. It is now in regular operation, producing images close to the diffraction limit in the near infrared, with uniform quality over a field of view of two square arcminutes. The present paper (I) is the first one in a two-paper review of GeMS. It describes the system, explains why and how it was built, discusses the design choices and trade-offs, and presents the main issues encountered during the course of the project. Finally, we briefly present the results of the system first light.
\end{abstract}


\section{Introduction}

Adaptive Optics (AO) is a technique that aims at compensating quickly varying optical aberrations to restore the angular resolution limit of an optical system. It uses a combination of Wave-Front Sensors (WFSs), to analyse the light wave aberrations, and phase correctors (e.g. deformable mirrors) to compensate them. In the early 1990s, astronomers experimented with the technique with the goal of overcoming the natural ``seeing'' frontier, the blurring of images imposed by atmospheric turbulence. See \cite{rousset1990first} for the results of the first astronomical AO system, {\sc Come-on}, and \cite{roddier1999adaptive} for a more general description of the first years of astronomical AO. The seeing restricts the angular resolution of ground-based telescopes to that achievable by a 10 to 50$\:$cm telescope (depending on the wavelength of the observation), an order of magnitude below the diffraction limit of 8-10$\:$m class telescopes. Two main limitations have reduced the usefulness of AO and its wide adoption by the astronomical community. First, the need for a bright guide star to measure the wave-front aberrations and second, the small field of view compensated around this guide star, typically a few tens of arcseconds. The first limitation was solved by creating artificial guide stars, using lasers tuned at 589$\:$nm, which excite sodium atoms located in the mesosphere around 90$\:$km altitude \citep{foy1985feasibility}. These Laser Guide Stars (LGS) could be created at arbitrary locations in the sky, thus solving the problem of scarcity of suitable guide stars. Nowadays, all of the major 8-10$\:$m ground-based telescopes are equipped with such lasers \citep{wizinowich2012progress}. The second limitation arises from the fact that the atmospheric turbulence is not concentrated within a single altitude layer but spread in a volume, typically the first 10$\:$km above sea level. Multi-Conjugate AO (MCAO) was proposed to solve this problem \citep{dicke1975phase,beckers1988increasing,ellerbroek1994first, johnston1994analysis}. Using two or more DMs optically conjugated to different altitudes, and several WFSs, combined with tomographic techniques, MCAO provides 10 to 20 times the field of view achievable with classical AO. MCAO as such was first demonstrated by MAD, a prototype built at the European Southern Observatory \citep{marchetti2007mad}. MAD demonstrated that MCAO worked as expected, but did not employ LGSs and as such could only observe a handful of science targets. 

GeMS is a Multi-Conjugate Adaptive Optics (MCAO) system in use at the Gemini South telescope. It uses five Laser Guide Stars (LGS) feeding five 16$\times$16 Shack-Hartmann Wave-Front Sensors (WFS), and needs three Natural Guide Stars (NGS) and associated NGS WFS to drive two Deformable Mirrors (DM). It delivers a uniform, close to diffraction-limited Near Infra-Red (NIR) image over an extended field of view of 2 square arcmin. GeMS is a facility instrument for the Gemini South (Chile) telescope, and as such is available for use by the extensive Gemini international community. It has been designed to feed two science instruments: GSAOI \citep{mcgregor2004gemini}, a 4k$\times$4k NIR imager covering $85''\times85''$, and Flamingos-2 \citep{elston2003performance}, a NIR multi-object spectrograph.  

GeMS began its on-sky commissioning in January 2011, and in December 2011, commissioning culminated in images with a FWHM of 80$\pm$2$\:$milliarcsec at 1.65$\:\mic$ (H-band) over the entire $85''$ GSAOI field of view.

This paper is the first in a two-paper review of GeMS. It makes extensive use of past published material to which the reader is referred. 

The plan of this paper follows chronologically the sequence of events in the history of GeMS to date. Section~\ref{sec:why_mcao} gives a general introduction about the subject of Multi-Conjugate Adaptive Optics (MCAO). Section~\ref{sec:history_of_gems} gives a short overview of how GeMS came to be. Section~\ref{sec:design_and_trade_offs} goes through the design, and discusses the various trade-offs that had to be made, due to cost or technological reasons. The next step after design and construction is the assembly, integration and tests (AIT); these are described in Section~\ref{sec:assembly_integration_and_tests} which also addresses in some detail the major issues that were encountered during the AITs. Finally, we present and discuss the results of the system first light. This paper is followed by paper II, which reports on GeMS commissioning, performance, operation on sky and upgrade plans.

\begin{figure}
  \caption[The LGS constellation] {The LGS constellation viewed from the side (about 100$\:$m off-axis) using a 500$\:$mm telephoto lens. The exposure time is 30 seconds.} 
  \begin{center}
  \includegraphics[width = 1.0\linewidth]{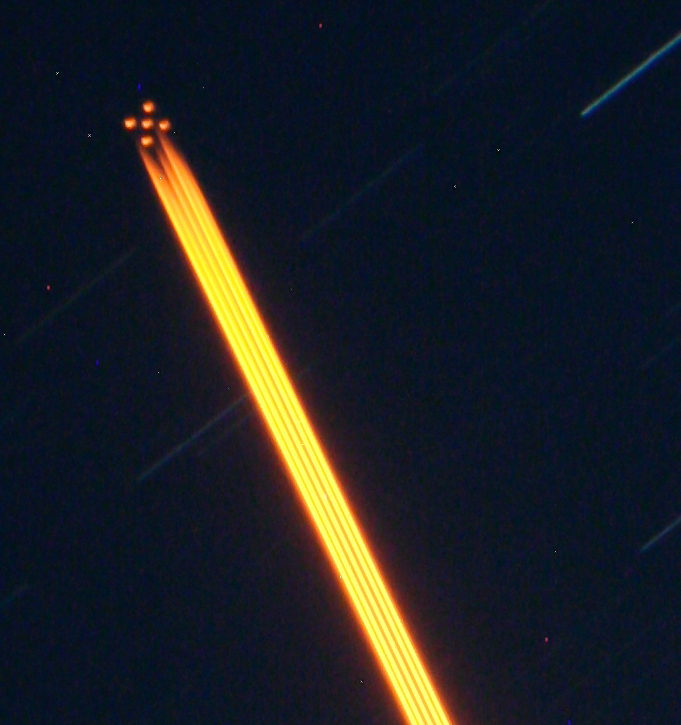}
  \end{center}
  \label{fig:lgs_constellation}
\end{figure}

\section{Why MCAO?} 
\label{sec:why_mcao}

In the late 1990s, LGS AO systems were emerging but still in their infancy. 
LGS AO held the promise of boosting the sky coverage accessible for AO compensation to very useful values (typically 30\% over the whole sky), but still suffered from two main limitations: anisoplanatism and focal anisoplanatism --- aka the cone effect. The cone effect \citep{tallon1990adaptive} is a consequence of the finite distance to the LGS ($\approx$ 90$\:$km above sea level). The effect of focal anisoplanatism on image quality depends on the turbulence Cn$^2$ profile and the diameter of the telescope. Under typical conditions on an 8$\:$m telescope, the Strehl ratio at 1.2$\:\mic$ is halved \citep{fried1994analysis}. In addition to the cone effect, angular anisoplanatism degrades the compensation quality when going off-axis from the LGS. 

MCAO uses several wave-front sensors (WFS) and deformable mirrors (DM) and tomographic-like wave-front reconstruction techniques to extend the correction off-axis, that is, obtain AO compensation not in a single direction, but over a field of view several times larger than the isoplanatic patch\footnote{The angle over which the error is lower than 1rd$^2$ is called the isoplanatic angle (or patch if one refers to the area). It varies from site to site and is wavelength-dependent. At Gemini South Cerro Pach{\'o}n it is about 20 arcsec at 1.6$\:\mic$.}. Thanks to the volumetric probing of the atmospheric turbulence and the tomographic processing, and compared to classical LGS AO, MCAO also virtually eliminates the cone effect \citep{rigaut2000principles}, increases the sky coverage and, by providing a significantly more uniform Point Spread Function (PSF), eases the astronomical data reduction process as well as improves the photometric and astrometric accuracy. MCAO was initially proposed by \cite{dicke1975phase} and then \cite{beckers1988increasing}, and the theory was developed by \cite{ellerbroek1994first}. The promises of MCAO attracted the interest of the science community \citep{ragazzoni2000adaptive,ellerbroek2000astronomy}, and around the year 2000, two projects started: GeMS (Gemini MCAO System) at the Gemini Observatory and MAD (Multi-conjugate Adaptive optics Demonstrator) at ESO \citep{marchetti2003mad,marchetti2007mad}.


\section{History of GeMS} 
\label{sec:history_of_gems}

Under the leadership of Ellerbroek (Project Manager) and Rigaut (Project Scientist), and once the kick-off effort had been approved by Matt Mountain (Gemini observatory director) and Fred Gillett (Gemini project scientist), GeMS passed a conceptual design review (CoDR) in May 2000. GeMS was to use three DMs, five LGSs and associated LGS WFSs, and three NGS WFSs (more details in section~\ref{sub:system_description}). It would consist of many subsystems (see section~\ref{sub:subsystems}); the main optical bench was to be attached to the telescope Instrument Support Structure (ISS), process the beam from the telescope and feed it back to the science instruments. From the beginning, a dedicated large near-infrared imager (that would become GSAOI) and a near-infrared multi-object spectrograph (Flamingos-2) were considered. A multi-IFU instrument was initially considered but rejected on the basis of cost and object density, which was too low for the Gemini 8$\:$m aperture over a 2 arcminutes field of view.

Following the CoDR, the Gemini science community was engaged during a three-day science case workshop in October 2000 at the CfAO headquarters in Santa Cruz. The workshop gathered 50 participants from the Gemini international astronomy community. Discussions focused on three main science themes: {\em ``Star formation and evolution in the Milky Way''}, {\em ``Nearby galaxies''} and {\em ``Distant galaxies''}, and eventually resulted in a document which once more emphasised the large gains GeMS would bring to existing programs and the new science it would enable \citep{rigaut2000science}. 

The team at Gemini also worked to advance theoretical knowledge specific to Multi-Conjugate AO \citep{flicker2000comparison,flicker2001sequence,flicker2002tilt,flicker2003methods}. 

As early as the CoDR, MCAO was recognised as the most challenging AO instrument ever built. It was relying on technology that was just starting to appear, and was pushing the limits on many fronts. One of the most challenging of these was the sodium laser. GeMS needed five LGSs. Gemini put together a comprehensive strategy to minimize the risks and cost of procuring the 50$\:$W guide star laser for GeMS \citep{dorgeville2000lgs}, including sodium layer monitoring campaigns that were needed to develop more informed requirements for the  lasers \citep{dorgeville2002gemini}.

After the CoDR, the project proceeded rapidly to PDR level, and a successful PDR took place in the Gemini North headquarters in May 2001 \citep{mcaopdr}.
The state of the project after the PDR is summarised in \cite{ellerbroek2003mcao}. Some time after the PDR, it was decided within Gemini to split up GeMS in subsystems, so as to get more tractable subcontracts. The prime motivation was to retain the Assembly Integration and Test (AIT) phase in-house --- an important step when considering the complexity of the system and the need for long term maintainability and upgrades. A number of subsystems were identified, that are listed below in section~\ref{sub:subsystems}. Consequently, there was no overall system Critical Design Review, but instead a whole new cycle of PDR/CDR by subsystem, held by the various vendors.



\section{Design and trade-offs} 
\label{sec:design_and_trade_offs}

\subsection{Simulations} 
\label{sub:simulations}

Initial simulations were carried out by two independent but similar software packages. LAOS, written by Ellerbroek \citep{ellerbroek2002wave,ellerbroek2002wave2} and \texttt{aosimul}, the precursor of YAO, written by Rigaut \citep{rigaut2013yao} are both Monte-Carlo physical image formation code that simulate atmospheric turbulence and AO systems, including WFS, DM, control laws, etc. Broadly speaking, these packages have similar functionalities. Results were cross-checked between the two packages, and bugs in the simulation codes were fixed. These simulations were extensively used as input for most of the design choices, such as the order of AO correction, the number and conjugate range of the deformable mirrors, and so on.

\subsection{Subsystems} 
\label{sub:subsystems}

Post PDR, a number of subsystems were identified:
\begin{itemize}
  \item The optical bench ({\sc Canopus}) including all opto-mechanics and the NGS WFS were built by Electro Optics System Technologies (EOST) \citep{james2003design}.
  \item The off-axis parabolas (two F/16 and one F/33) were polished by the Optical Science Center at University of Arizona.
  \item The LGS WFS assembly was made by the Optical Science Company (tOSC). It has five arms. As in any Shack-Hartmann WFS, there were stringent requirements in pupil distortion and wave-front quality. The particular challenge was to keep distortions and aberrations low for {\em off-axis} LGS over a wide LGS range (80 to 200$\:$km). Rob Dueck at tOSC went through 7 iterations for the optical design, to end up with a solution with 8 optics per LGS channel (4 are common to the 5 paths, 4 are independent per path) and 8 actuated stages for the whole assembly.
  \item The Real-Time Computer was also built by tOSC. It uses a dedicated, OS-less TigerShark DSP platform (2$\times$6 DSPs), with a windows host computer (communications, GUI and interfacing with the DSP dedicated PCI bus).
  \item The 3 DMs, DM0, DM4.5 and DM9 (the number refers to their conjugation altitude) were built by CILAS. An important note on the number of DMs: GeMS was designed and integrated with 3 DMs. However, following issues with one of them (see section~\ref{ssub:dm0_actuator_failing}), the system has been working with only 2 DMs --- the ground and the 9$\:$km DMs --- for most of the commissioning. The intermediate DM will eventually find its way back into {\sc Canopus}. This is why the reader will find throughout this paper sometimes confusing references to the system in both its 2 and 3 DM configurations.
  \item The DM electronics were built by Cambridge Innovations. CILAS DMs take $\pm$ 400$\:$V and the phase delay induced by the electronics had to be small at the loop maximum rate of 800$\:$Hz.
  \item The Beam Transfer Optics, because of their very tight integration with the telescope and observatory operations were designed and built in-house at Gemini. 
  \item The Laser Launch Telescope (LLT) was build by EOST. LLT are generally considered non-challenging subsystems and too often are not given enough attention. As a result, they often fail, or fall short of the original specifications. Challenges of this subsystem are optical quality, flexure and, more importantly, athermalisation to avoid focus drifts in the course of an observing night. Focus drift would result in LGS spots FWHM degradation, which is difficult to measure as they have the same signature than, say, seeing or laser beam size degradation.
  \item The laser was built by what was initially Coherent Technologies Inc,   which turned into Lockheed Martin Coherent Technologies (LMCT) shortly after the contract was signed. Although there were some discussions initially whether it was better to go for five 10$\:$W lasers or one 50$\:$W laser, it soon appeared than even if developments were more challenging, the latter solution was preferable for cost, space and maintenance reasons. Many more details can be found in \cite{dorgeville2002gemini}, \cite{hankla2006twenty}, and \cite{dorgeville2003laser}
\end{itemize}

Of paramount importance were the software, the safety systems and the management. The software represented a very significant effort. Functionalities to be provided went from low-level control of hardware (e.g. BTO motors), to adapting the Gemini observation preparation tool for use with GeMS, through the real-time code, the AO simulation, the AO real-time display and diagnostic tool, airplane detection code, satellite avoidance, laser traffic control, etc. Some elements can be found in \cite{boyer2002gemini}, \cite{bec2008ascam}, \cite{dorgeville2012gemini} and \cite{trancho2008gemini}.

The management and systems engineering were done in-house. GeMS has had four project managers in the 13 years span of the project to date; Brent Ellerbroek, Mike Sheehan, Maxime Boccas and Gustavo Arriagada. \cite{boccas2008gems} exposes in some details management issues, schedule and resources. 


\subsection{Sodium monitoring campaigns} 
\label{sub:sodium_monitoring_campaigns}

To be able to make an informed decision about the laser power requirements, the design team realised early that there was a need for sodium layer characterisation at or close to Cerro Pach{\'o}n. A site monitoring campaign was set up at the AURA-operated Cerro-Tololo Interamerican Observatory (CTIO) in Chile in 2001 and 2002. It comprised five observation runs of typically 10 nights each, strategically scheduled every 3 months to get a proper seasonal coverage. Both the CTIO 1.5$\:$m and Schmidt telescopes were used. The goal was to measure sodium layer profile, and derive sodium density, layer altitude and structure on a minute time scale. The set-up and results are described in \cite{dorgeville2003preliminary} and \cite{neichel2013sodium}. The laser was a dye laser on loan from Chris Dainty's group at the Blackett Laboratory, Imperial College, London. The power propagated on sky was in the 100-200$\:$mW range. Results confirmed seasonal variations, with a sodium column density minimum around $2\times10^9 \; {\rm atoms/cm}^{2}$ occurring in the southern hemisphere summer. The GeMS instrument design and the laser requirements were based on this rather conservative value, as summer is the best season to observe, given that statistically speaking it offers better seeing and clearer weather conditions. These data also provided useful information on the sodium layer altitude variations; an important quantity when considering how often the focus information should be updated on a LGS system, using a reference Natural Guide Star --- typically, but not necessarily, the same NGS as for Tip-Tilt (TT). It was found by \cite{dorgeville2003preliminary} that the RMS variation of the sodium layer mean altitude is on the order of 15$\:$nm over 30 seconds, thus that an integration time of 10 seconds on the NGS focus WFS would be adequate.


\begin{table}
\caption{GeMS in numbers. For acronyms see Section~\ref{ssub:canopus_the_optical_bench}.}
\begin{tabularx}{0.475\textwidth}{ll}
\hline \hline
DM conjugate range      & 0, 4.5 and 9$\:$km \\
DM order            & 16, 16 and 8 across the 8$\:$m beam \\
Active actuators    & 240, 324, 120  (total 684) \\
Slaved actuators    & 53, 92, 88 (total 233)\\ \hline
5 LGS WFS           & SHWFS, 16$\times$16 (204 subap)\\ 
                    & 2$\times$2 pix/subap., $1.38''$/pixel \\ 
3 NGS TT WFS        & APD-based quadcells, $1.4''$ FoV \\ 
1 NGS Focus WFS     & SHWFS, 2$\times$2 \\ 
                    & Light split with TT WFS \#3 \\
WFS sampling rate   & Up to 800$\:$Hz \\ \hline
TT WFS magnitudes   & 3$\times$R=16 (actual, spec was 18)  \\ 
                    & for 50\% Strehl loss in H \\ \hline 
LGS const. geometry        & (0,0) and ($\pm$30,$\pm$30) arcsec \\ 
Launch Telescope    & Behind telescope M2, 45$\:$cm $\oslash$ \\ \hline
Wave-front control& Minimum Variance Reconstructor \\
 & Decoupled LGS/NGS control\\
\hline \hline
\end{tabularx}
\label{tab:system_description}
\end{table}

\begin{figure*}
  \caption[GeMS synaptic diagram] {GeMS ``synaptic'' diagram, showing major subsystems and their relationship, all loops and offloads.}
  \begin{center}
  \includegraphics[width = 1.0\linewidth]{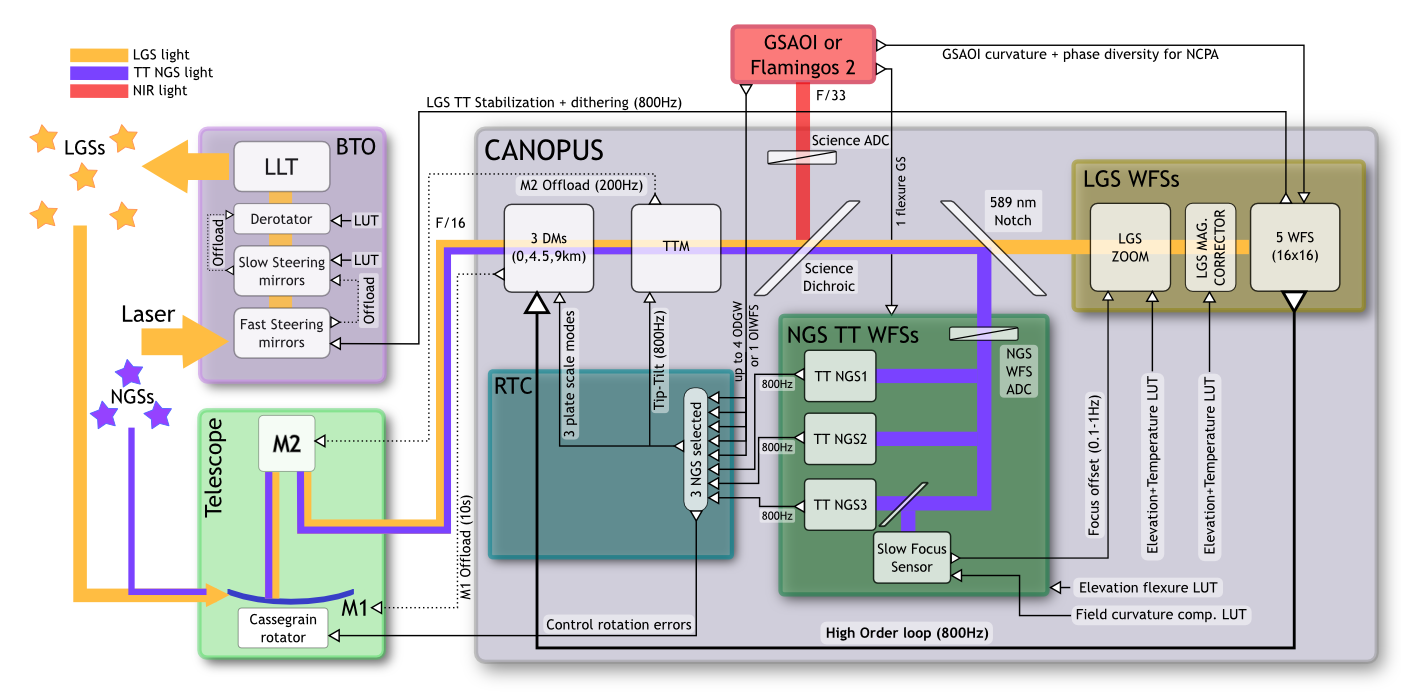}
  \end{center}
  \label{fig:gems_synaptic_diagram}
\end{figure*}

\subsection{System Description} 
\label{sub:system_description}

Table~\ref{tab:system_description} presents a top-level description of the main system components. Figure~\ref{fig:gems_synaptic_diagram} shows a diagram of the major subsystems and their inter-connections, including the many loops and offloads. These are discussed in more details below.

\subsubsection{Laser and Laser Guide Star control} 
\label{ssub:laser_and_laser_guide_star_control}

GeMS works with a Laser Guide Star (LGS) constellation (see Figure~\ref{fig:lgs_constellation}) resembling the 5 spots on the face of a die: 4 of the LGS spots are at the corners of a 60 arcsec square, with the 5\textsuperscript{th} positioned in the centre. These LGSs are produced by a 50$\:$W laser split into 5 distinct 10$\:$W beacons by a series of beamsplitters. The on-sky performance of the LGSF is described in \cite{dorgeville2012gemini}. The laser bench and its electronics enclosure are housed inside a Laser Service Enclosure (LSE), located on an extension of the telescope elevation platform (a Nasmyth focus for other telescopes). The Beam Transfer Optics (BTO), a subsystem of the LGSF, relays the laser beam(s) from the output of the Laser system to the input of the Laser Launch Telescope (LLT) located behind the telescope secondary mirror. Beside relaying the laser light from the laser to the LLT, the BTO ensures the slow and fast compensation of telescope flexures and constellation alignment control. Figure~\ref{fig:gems_synaptic_diagram} shows a diagram of the main BTO elements and their interactions. Because Gemini South is an Alt-Az telescope and because the LGSs are launched from a small telescope fixed on the back of the secondary mirror, the laser constellation must follow the telescope field rotation and de-rotate to keep the LGS spots fixed with respect to the AO bench. This is achieved by a K-Mirror (KM) located in the BTO. Alignment and control of each LGS in the constellation is provided by five fast Tip-Tilt platforms, called FSA (for Fast Steering Array). The FSA platforms offload average Tip-Tilt to a Centering and Pointing Mirror (CM and PM). The averaged rotation accumulated on the FSA platform is also offloaded to the K-mirror.

\subsubsection{Laser safety systems} 
\label{ssub:laser_safety_systems}

Operating and propagating guide star lasers is delicate. These are Class IV lasers which have very well defined and stringent safety regulations (for good reasons). As far as propagation is concerned, when MCAO operation was first discussed, there was good experience from a few other facilities: namely, the Shane 3$\:$m telescope at Lick Observatory, the Starfire Optical Range in New Mexico and the Keck II telescope on Mauna Kea. On U.S. soil, propagating guide star lasers requires approval from the Federal Aviation Administration (or the local equivalent outside the U.S.), and using a secure and approved system to avoid propagating in the direction of planes; something that was ---and still is--- performed at Gemini by human spotters. Depending on the local authorities, alternative solutions have been sought, that may involve arranging for a no-fly zone (ideal) or using automated wide field or thermal cameras to detect air planes and automatically stop laser propagation. In the US, it has been historically challenging to obtain authorisation from the FAA to replace human spotters by automated systems. However, Gemini gave it a try: Hardware was procured, and significant development efforts were dedicated to write plane detection software \citep{bec2008ascam}, with good success. Planes were generally detected during test runs, from ten degrees elevation up (GeMS can not be used at elevation lower than 40 degrees, so this leaves some margin), with very high probability\footnote{Note that this is the key word, and ``very high'' probability may actually not cut it}. This effort was however cut short before the software was truly a finished product, as it appeared that it would be challenging to obtain approval from the FAA (at Gemini North, or its Chilean equivalent, the Direcci{\'o}n General de Aeron{\'a}utica Civil, at Gemini South). 

Another agency with which GeMS operations have to coordinate with is the US Laser Clearing House (LCH). This agency coordinates high power laser upward propagation to avoid hitting space assets; e.g. satellites which stabilisation sensors could be disturbed or potentially damaged by the laser light. List of targets have to be sent by the observatory to the LCH, which returns a list of time windows during which propagation is or is not authorised. There are several levels of security (both automated and human) at Gemini during observing to prevent propagating during a LCH no-propagation window. Generally, but not always, the observing plan for the night is put together such as to avoid long no-propagation windows, by the proper selection of targets (no-propagation windows are on a target basis). See \cite{dorgeville2008gemini}, \cite{dorgeville2012gemini} and \cite{rigaut2005practical} for more details on all the laser safety systems.

A third and final concern when propagating lasers is interference with neighbour facilities. Rayleigh scattering of the 589$\:$nm light (or whatever other wavelength in the case of a Rayleigh LGS) can definitely wreak havoc on images or spectra from telescopes situated up to a few kilometres away, so coordination with neighbour facilities --- and possibly the establishment of policies --- are a must (e.g. should priority be given to the first telescope on a given target or to non-laser telescopes?). The software, initially written by Keck for laser traffic control at Mauna Kea (LTCS, \cite{summers2012decade}) was adapted for operation at Gemini South. Currently, the only neighbour telescope is SOAR; Studies were done and measurement taken and it was concluded that CTIO (a distance of 10$\:$km as the crow flies) was not affected by Gemini's laser.


\begin{figure}
  \caption[View of {\sc Canopus}] {View of the AO optical bench, {\sc Canopus}}
  \begin{center}
  \includegraphics[width = 1.0\linewidth]{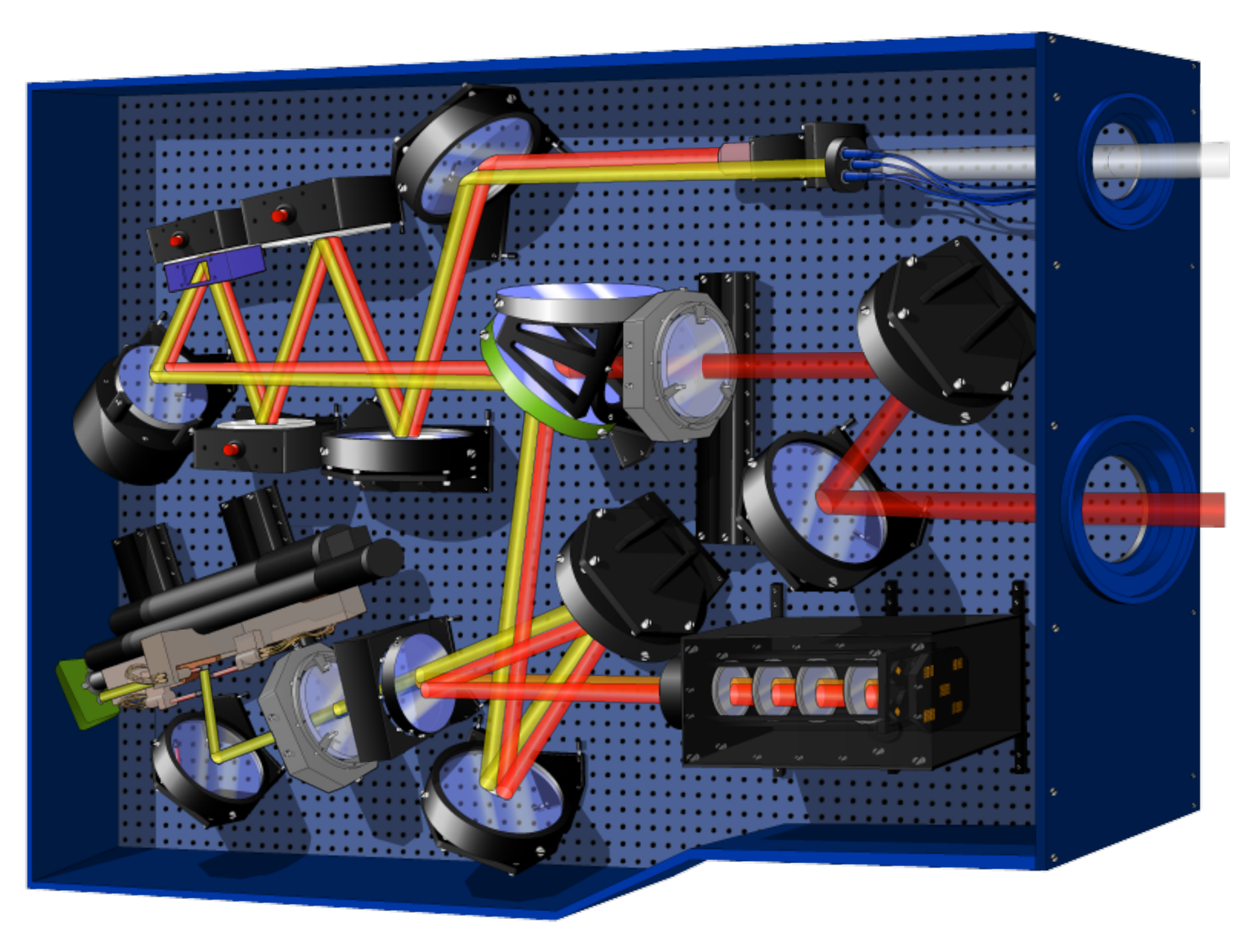}
  \end{center}
  \label{fig:gems_diagram}
\end{figure}


\subsubsection{{\sc Canopus}, the optical bench} 
\label{ssub:canopus_the_optical_bench}

The optical design was done by Richard Buchroeder \citep{james2003design,bec2008gemini}. It is a plane design, intended to simplify alignment and maintenance. Figure~\ref{fig:gems_diagram} shows the (vertical) AO optical bench, {\sc Canopus}, attached to the Gemini Cassegrain-located Instrument Support Structure (ISS). Through the ISS, the Gemini telescope F/16 beam is re-directed to the {\sc Canopus} bench via the flat AO-fold mirror. The MCAO correction is performed by three DMs conjugated to 0, 4.5 and 9 km (hereafter called DM0, DM4.5 and DM9 respectively) and one Tip-Tilt Mirror (TTM). Following the DMs, a first dichroic beam splitter is responsible for separating visible from NIR light, sending the former to the WFSs, and the latter to the science output with a F/33.2 focal-ratio to feed the instruments. The visible light directed toward the WFS is split into a narrow range around 589$\:$nm to illuminate the five LGS Wave-Front Sensors (LGSWFS); the remaining visible light goes to the Natural Guide Star WFS (NGSWFS). Figure~\ref{fig:gems_synaptic_diagram} provide a complementary, functional view of the entire GeMS system. More details can be found in \cite{bec2008gemini}.

The whole optical bench is ``sandwiched'' on either side by electronic enclosures that house all the control electronics for mechanical stages and calibration sources, as well as the Real-Time Computer, the DM high voltage power supplies, the TT mirror control electronics, the Avalanche Photo-Diodes (APD) counters and the CCD controllers. The entire instrument weighs approximately 1200$\:$kg and fits in a 2$\times$2$\times$3$\:$m volume. 


\subsubsection{LGSWFS and LGS-related loops} 
\label{ssub:lgswfs_and_lgs_related_loops}

The LGSWFS is composed of five 16$\times$16-subaperture Shack-Hartmann. All five LGSWFS are identical, except for their pointing. The LGSWFS pixel size if $1.38''$ and each subaperture is sampled by 2$\times$2 pixels (quadcell configuration). The LGSWFS assembly contains 8 stepper mechanisms (two zoom lenses and six magnificators) used to accommodate for the changes in range of the LGSs (change in telescope elevation or changes in the Na layer altitude), as well as to compensate for flexure and temperature variations present at the ground level. The current range accessible with the LGSWFS is from 87.5$\:$km to 140$\:$km, corresponding to an elevation range of 90 to 40 degrees respectively. The following parameters need to be controlled: (i) the DM0 to each LGS WFS registration, (ii) the WFS magnification and (iii) the focus phase errors. These are controlled using look up tables (LUT) that depend on elevation, Cassegrain rotator position and temperature \citep{neichel2012identification}.

The LGSWFS provides a total of 2040 slope measurements from 204 valid subapertures per WFS. The use of quadcells require the knowledge of a calibration factor, the centroid gain, to transform the quad cell signal (unitless) into a meaningful quantity, e.g. the spot displacement in arcsec. This centroid gain is proportional to the size and shape of the LGSWFS spot, which changes with laser beam quality, seeing and optical distortions. The calibration of the LGSWFS centroid gains is thus done in soft real-time, by a procedure described in \cite{rigaut2012gems}.

The Tip-Tilt signal from each of the LGSWFS is averaged and sent to the BTO-FSA to compensate for the uplink laser jitter, and keep the laser spots centred at a rate of up to 800 Hz. The remaining modes are used to compute the MCAO high-order correction applied at a rate of up to 800 Hz by the three DMs. The total number of actuators is 917 including 684 valid (seen by the WFSs) and 233 extrapolated \citep{neichel2010gemini}. Unsensed actuators are very important for AO systems with a DM conjugate to an altitude higher than the ground since they affect science targets located in the outskirts of the field of view. The phase reconstruction and DM voltage control is done by a Real Time Computer (RTC). The reconstruction algorithm is described in \cite{neichel2010gemini}. The RTC also computes the averaged first 12 Zernike modes on DM0 and offloads them to the primary mirror of the telescope at a lower rate. 


\subsubsection{NGSWFS and NGS-related loops} 
\label{ssub:ngswfs_and_ngs_related_loops}

The NGSWFS consists of three probes, each containing a reflective pyramid that acts like a quadcell feeding a set of 4 fibers and corresponding Avalanche Photo Diodes (APDs). The three probes can be placed independently within a 2 arcmin acquisition field. Each probe provides a tip and a tilt measurement at a rate of up to 800$\:$Hz (capped by the LGS WFS rate). The weighted averaged signal over the probes gives the overall Tip-Tilt and is used to control the TTM. The weights depend on the noise and location of the WFS.

The TTM offloads its average pointing to the secondary mirror of the telescope at a rate of up to 200$\:$Hz. A rotation mode is estimated from the probe positions, and offloaded to the instrument rotator. Finally, the differential Tip-Tilt errors between the three probes are used to control the plate scale modes (also called Tilt-Anisoplanatic or TA modes \citep{flicker2002tilt}). The plate scale errors are compensated by applying quadratic modes with opposite signs on both DM0 and DM9. The reconstruction algorithm follows the scheme  described in \citet{neichel2010gemini}. As there is no offloading possibilities for DM9, the position of the probes in front of their respective guide star must be optimised. This is done during acquisition when the Tip-Tilt errors are averaged over a 10$\:$s period, and each of the NGS guide probes is moved in order to lower this error below a given threshold. After setting-up on an object, the individual probes are locked on a common platform, fixing the relative distance between them. During an observation, only the common platform moves, hence conserving the image plate scale and allowing for astrometry measurements.

One of the probes contains a small beam splitter that sends 30\% of the light to a Slow Focus Sensor (SFS). As the LGSs are used to compensate for atmospheric focus, any changes in the sodium layer altitude cannot be disentangled from atmospheric focus changes. To cope with this effect, the focus on a NGS is monitored by the SFS. The SFS is a 2$\times$2 Shack-Hartmann and the focus error it measures is sent to the LGSWFS zoom to track the best focus position as seen by the science path. The SFS control strategy is described in \citet{neichel2012science}.

To compensate for potential differential flexures between the AO bench and the instrument, a flexure loop uses the signal coming from an On-Instrument (OI) WFS on the science instrument. The flexure signal is used to drive the position of the NGS WFSs with an update rate between 1 and 30$\:$s.



\subsubsection{Control} 
\label{sub:control}

The RTC is responsible for measuring and correcting wave-front errors. It was built by Stephen Brown at tOSC and is described in \cite{bec2008gemini}. The signal from the five laser guide star wave-front sensors and three natural guide star wave-front sensors is collected and analysed to control the three deformable mirrors and the tip/tilt mirror. 

The RTC was built using off the shelf components. A Pentium CPU hosts the graphical user interface and runs miscellaneous background tasks. The host implements the TCP/IP layer to the observatory command and status interface. Hard real time computations and control of the hardware (5 LGSWFS, 3 DMs, 3 NGS TT WFS and the TTM) are handled by an array of 12 TigerSHARCs DSPs (two TS201S cards hosting six 550MHz DSPs each) mounted on a PCI extension chassis.  Distribution of tasks on different DSPs allows a high degree of parallelism. Communication between the different processes in the RTC is accomplished using shared memory. Different ring buffers store real-time information. That content of the buffer can be saved on disk to be accessible to background optimisation processes and diagnostic utilities. Stringent operations were implemented in assembly code to meet the high throughput and low latency requirements of GeMS. The overall latency (last received pixel to last command sent to the DM power supply) was measured at approximately 50$\mu{\rm s}$ (see Sect. \ref{ssub:rtc}).

\begin{figure}
  \caption[GeMS Real-Time Display and Diagnostics] {Snapshot of the Real-Time Display \& Diagnostics in action}
  \begin{center}
  \includegraphics[width = 1.0\linewidth]{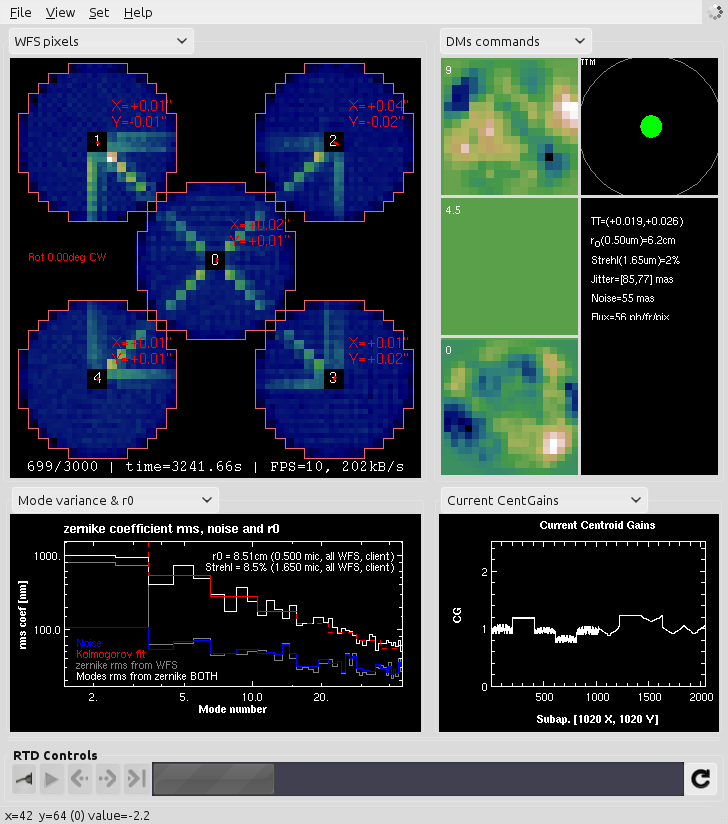}
  \end{center}
  \label{fig:rtd}
\end{figure}

The LGS control law implements a leaky integrator of the form
\begin{equation}
y[n] = (1-l)\;y[n-1]+g\;e[n],
\end{equation}
where $y[n]$ is the command at time $n$, and $e[n]$ is the wave-front error computed from the WFS slopes. The integrator loop gain is $g$ and $l$ is a leak term required to reduce the effect of poorly sensed modes. The leak turned out to be of utmost importance during the integration and test phase, to be able to work in closed loop even when the system was not perfectly aligned.

The tip-tilt control law is given by:
\begin{equation}
y[n] = b_1 \; y[n-1] + b_2 \; y[n-2] + a_1 \; e[n] + a_2 \; e[n-1],
\end{equation}
where $a_1$, $a_2$, $b_1$ and $b_2$ are coefficients that can be set to reduce the tip-tilt error. Some more  complex control laws (Kalman, H$_2$, H$_{\infty}$) have also been tested for vibration suppression \citep{guesalaga2012design,guesalaga2013performance}, and may be implemented for operation in the future (see paper II for more details). 

GeMS reconstructors were originally generated based on synthetic interaction matrices. \cite{rigaut2010sample} lists advantages and drawbacks of this choice. The current scheme is now based on measured experimental interaction matrices. Note that the interaction matrix depends on the zenith angle due to the changing range to the laser guide stars. A regularised inverse of the interaction matrix is used to reconstruct the wave-front.

Since the RTC hardware was not specified to perform two matrix multiplies, pseudo open-loop control is not possible. Therefore, true minimum-variance reconstructors cannot be implemented.


\subsection{Software} 
\label{sub:software}

This section focuses on the high level software associated with {\sc Canopus} operation. For information concerning the BTO, Laser and laser safety systems software, the reader is referred to \cite{dorgeville2012gemini,dorgeville2008gemini}.

\subsubsection{High level control and display} 
\label{ssub:myst}

{\sc Myst} is the top-level engineering graphical user interface for the operation of {\sc Canopus}. It has been described extensively by \cite{rigaut2010myst}. As a GUI, {\sc Myst} essentially fulfils two functions: it provides convenient control of the {\sc Canopus} functionalities (loop control, mechanism control, control matrix creation, etc) and a Real-Time Display and Diagnostic (RTDD) tool. The RTDD can display raw (WFS slopes, DM actuators, etc) and processed (e.g. DM projection on Zernike modes, $r_0$ estimation by fitting of Zernike mode variance) information at 10-20$\:$Hz. Figure~\ref{fig:rtd} gives an example of what the RTDD looks like (pull down menus allow independent configuration of each graphical pane). 

\subsubsection{Off-line packages: data reduction and calibrations} 
\label{ssub:off_line_packages_data_reduction_and_calibrations}

In addition to MYST, a number of high level software packages were developed for off-line data analysis or calibration. 
\begin{itemize}
  \item {\bf WAY} is a generic wavefront reconstruction and display tool and was used with the 24x24 diagnostics Shack-Hartmann wavefront sensor used during the AIT \citep{garcia2010calibration}.
  \item {\bf OPRA} \citep{rigaut2011gems,gratadour2011tomographic} is a phase diversity package that uses the new tomographic method described in section~\ref{ssub:opra} . It is used regularly since the beginning of the commissioning (needs both {\sc Canopus} and GSAOI) to null non-common path aberrations. 
  \item {\bf YAO} \citep{rigautyao,rigaut2013yao} is a software package and library to simulate AO systems. It is derived and expanded from \texttt{aosimul} (See Section~\ref{sub:simulations}). It was used extensively, as the library that power other tools (MYST,WAY), as a simulator to interface with the other software tools for testing, and finally and most importantly, to generate synthetic control matrices for the system when this method was in used. 
  \item {\bf ASCAM} \citep{bec2008ascam} is the software package developed and tested at Gemini for the detection of moving objects (also works for UFOs), as referred to in section~\ref{ssub:laser_safety_systems}.
\end{itemize}

All of these packages make use of either python, C/C++, yorick, or a combination of these.

\subsubsection{Low level software} 
\label{ssub:low_level_software}

Low level software include the RTC code \citep{bec2008gemini}, the {\sc Canopus} mechanisms control code, the BTO code and the laser safety code \citep{dorgeville2012gemini}. The RTC has been described in Section~\ref{sub:control}. Control of essentially all the motorized stages and status information from a variety of sensors mounted on {\sc Canopus} and the BTO is implemented using the Experimental Physics and Industrial Control System \citep{dalesio1994experimental}. EPICS is a standard framework adopted at Gemini. It provides low-level drivers to control hardware (motor controllers, digital and analog input/output etc.), a network transparent layer (Channel Access) to distribute command and status information, a variety of graphical user interfaces builder to generate high level applications. Approximately half of this software was done in-house. The other half was contracted out to the UK-based company Observatory Science Limited.



\subsection{Design choices and trade-offs} 
\label{sub:design_choices_and_trade_off}

During the design phase, a number of choices and trade-offs had to be made; either specific to AO, LGSs, or the multi-LGS, multi-conjugate aspect. Because GeMS is still to date the only LGS MCAO system ever built, it is interesting to comment further on these trade-offs.

\begin{itemize}
  \item {\bf Reflective design}: This one is easy enough. As for many AO systems, because of the wide wavelength range (from 450$\:$nm to 2.5$\:\mic$), a refractive design for the common path optics would have been very challenging, especially in terms of optical throughput (chromatism correction and coatings).

  \item {\bf Two output F-ratios}: There was an intense debate about the choice of F/33 for the science output. This was going against the philosophy of Gemini in which the AO was just an adapter and should deliver the exact same F/16 beam as the telescope to the science instruments (as Altair was for instance doing at Gemini North). Eventually, it was recognised that going for F/33 would lower the risk of non-common path aberrations (smaller optics) and actually fit better existing instruments (provide twice the plate scale) and would make the design of AO-dedicated instruments like GSAOI simpler.

  \item {\bf Lower actuator density on DM9}: In MCAO, because of the field of view, the area to be controlled is larger on high altitude DMs/optics than in the pupil. For instance, DM0, conjugated to the telescope pupil, has to ``control'' an equivalent area of 8$\:$m in diameter. Now when getting to 9$\:$km, the equivalent area (for the GeMS field of 2 arcmin) becomes $8 + 2\times60 \times 9\:10^3 \times 4.8\:10^{-6} = 13.2\:$m in diameter. This means $(13.2/8)^2=2.7$ times the area, for a slab of turbulence that contains a much smaller fraction of the turbulence that what is at the pupil/ground. For cost and complexity reasons, it was thus decided to double the actuator pitch on the highest DM. This should still deliver adequate phase variance reduction, provided of course that the outer scale of turbulence in this high altitude slab was of the same order as it is in the ground layer, something that was unknown at the time (and still is).

  \item {\bf Field of view}: The MCAO science case clearly showed that the largest possible field of view ($>$ 1$\:$arcmin) was of the utmost importance for the majority of the science cases. This pushed toward large a field of view. Factors that limited the design ambitions included (a) the feasibility and cost of the GSAOI detector array (4k$\times$4k, in a 4 Hawaii-2RG detector package), given the need to approximately sample at J band ($\approx$20$\:$milliarcsec/pixel), and (b) the fact that the ISS, as built, only transferred a 2 arcminute field through the AO port. The latter constraint would have been very difficult and costly to remedy. Finally, the design team adopted a 2 arcminutes maximum field of view, with a central 85$\times$85 arcseconds over which the compensation quality would be maximised.

  \item {\bf Tip-Tilt mirror not at the pupil plane}: The TT mirror (TTM) is conjugated at about 3$\:$km below ground. The conjugated altitude is not by itself a problem (TT will be corrected the same whatever the conjugation altitude is), but the fact that it is located in-between the DMs and the WFS implies that when it tilts, the TTM will move the image of the DM as seen by the Shack-Hartmann WFS, and thus modify the DM-to-lenslet registration. However, the induced misregistration for e.g. DM0 is only 4\% of a subaperture per arcsecond of TTM motion, and was deemed acceptable considering the added complexity, cost, and loss in throughput of an optical design where the pupil would have been re-imaged on the TTM.

  \item {\bf Laser launch behind M2}: In GeMS, there is only one Laser Launch Telescope (LLT) to launch the five LGS beams, and it is located behind M2. The drawback is that in this configuration, any WFS sees the Rayleigh scattering from the other LGS beams (see for instance the upper left WFS display in Fig~\ref{fig:rtd} and its X pattern of Rayleigh illuminated subapertures). The alternative of using side launched lasers was not seriously considered for the following reasons. It was believed that image elongation resulting from a side projection was very bad in term of SNR (we know differently since then \citep{thomas2008study,robert2010shack}). Also, entirely avoiding Rayleigh backscatter would have meant (a) getting rid of the central LGS, which at the time was deemed to improve significantly the uniformity across the field of view and (b) using four LLTs, which would have meant a large increase in cost and operation reliability. At the time, it was also believed, due to the absence of any measurements, that the fratricide could be subtracted away from the affected subapertures. This is not feasible, in part because of aerosol and fast laser power variability, but primarily because the Beam Transfer Optics uplink pointing adjustment mirrors are not in a pupil plane, causing the Rayleigh (near field) to move when the LGS pointing changes \citep{neichel2011sodium}. A last factor was a lower sodium coupling coefficient than initially expected (between the laser and sodium atoms), which reduced the ratio between the LGS and the Rayleigh components. Rayleigh backscatter disables about 20\% of the subapertures, but the impact on performance has not been well studied.

  \item {\bf Number of LGS}: This was based on an exhaustive study using both Ellerbroek \citep{ellerbroek2002wave,ellerbroek2002wave2} and Rigaut \citep{rigautyao} simulation codes, and driven by both Strehl and Strehl uniformity requirements. In retrospect, it may have been preferable to break the redundancy in the constellation geometry, to avoid or reduce the appearance of invisible modes. Although these do not appear to be much of a problem in the current two DM configuration (see section~\ref{sub:issues}), it could become so when DM4.5 makes its way back into {\sc Canopus}.

  \item {\bf Number of NGS}: Alternative schemes to compensate for anisoplanatism TT modes (TT but also plate scale) were looked into \citep{ellerbroek2001methods}, for instance using a combination of Sodium and Rayleigh laser beacons. Eventually, the simplest method of using three NGS was retained. Detailed sky coverage evaluation showed that the need for three NGS --- compared to one for LGS AO --- is not detrimental, as the three TT NGS can be located much further away (up to 60 arcsec) than the single TT NGS in LGS AO (up to 30 arcsec typically) for comparable performance \citep[e.g.][]{fusco2006sky}.

  \item {\bf Quadcells}: At the time GeMS was designed, the detector market looked a lot different than it does right now. 
  To fit 16$\times$16 Shack-Hartmann spots, the best available low noise detectors were the EEV CCD39, with 3.5 electrons read noise. These CCDs have 80$\times$80 pixels, so could only fit 4$\times$4 pixels per subapertures. It was thus decided to go for quadcells (2$\times$2 pixels) in each subaperture, and keep one guard pixel on each side to avoid cross illumination between subapertures (combined with a field stop). Quad cells come with a lot of issues though: pixel edge diffusion degrades the FWHM and thus the SNR (section~\ref{ssub:wfs_ccd_pixel_modulation_transfer_function}); but the main issue is centroid gain calibrations. The centroid gain is the constant of proportionality between the quad cell measurement and the physical spot displacement. It is also a function of the spot size and shape. During operation, because of changes in seeing, in laser beam quality, and in the Sodium layer thickness/profile, the spot size and thus the centroid gain will change. An inordinate amount of effort was devoted to centroid gain calibration and/or issues related to centroid gains. If there is one lesson learned from GeMS, it is this one: Don't use quad cells in a SHWFS if you can avoid it.

\item {\bf DM altitude conjugation}: The DM altitude conjugation choice was made based on initial numerical simulations. In particular, there was some debate regarding the number of DMs (two vs three). For the given field of view and targeted wavelengths, the three DM configuration was found to be significantly more robust to changes in the Cn2 profile and was finally adopted. Based on more recent measurements derived from GeMS itself, \citep{cortes2013analysis} it appears that 9$\:$km is too low to compensate for the turbulence induced by the jet-stream, usually located between 11 and 12$\:$km. This is particularly impacting performance when the telescope is pointing at low elevation and the apparent distance to the jet stream is larger.

\end{itemize}




\section{Assembly Integration and Tests} 
\label{sec:assembly_integration_and_tests}

Assembly, Integration and Tests (AIT) of {\sc Canopus} took place at the Gemini south base facility, in La Serena, Chile. The first elements were received in 2007, and integration was completed by the end of 2010, when {\sc Canopus} was sent to the telescope. All the sub-systems were assembled and tested in the lab during that period. No formal and overall Acceptance Test was performed before sending the instrument to the telescope. A good overview of the activities and performance of the system can be found in \cite{boccas2008gems}, \cite{neichel2010gemini} and \cite{garcia2010calibration}. Below we summarize the main results obtained during this {\sc Canopus} AIT period.

\subsection{Beam Transfer Optics}
\label{sub:btoAIT}

\subsubsection{Optics}
\label{ssub:btooptics}

Construction and integration of the BTO started in 2007, and finished in summer 2010.
Integration of the BTO optomechanics on the telescope, with its 32 optics and 26 motors, took a very significant amount of resources. Given its tight integration with the telescope, telescope access time also turned out to be an issue, as it was competing with maintenance, day time instrument calibrations, etc.

The LLT was installed on the telescope in late 2007 \citep{dorgeville2008gemini}, and first optical quality measurement were done on-sky soon after. The GS BTO throughput was measured to be on the order of 60\%. This is under the original specification of 75\%, and was attributed in large part to suboptimal BTO coating specifications that failed to take into account proper polarisation control considerations.

\subsubsection{Mechanics}
\label{ssub:btomech}

Inelastic flexures, probably due to the long length of the BTO, are preventing the use of only LUT to keep the alignment. Active control based on an optical feedback from pre-alignment cameras, and the laser pointing on the sky (see Sect. \ref{ssub:laser_and_laser_guide_star_control}) is mandatory to keep the beams perfectly aligned all along the long BTO optical path. 
The original BTO design also included a fast Laser Beam Stabilisation system (LBBS) that would compensate for vibrations and fast laser beam drifts in the BTO while propagating at full power. It appeared that this real-time stabilisation was not required and that only the remote re-alignment mentioned above was enough to keep the alignment on the BTO. Finally, the main issue with the mechanical performance of the BTO was related to the mount of the Fast Steering Array (FSA) mirrors. The original mechanical design included a clamping of the piezo body of these Tip-Tilt platforms, causing an accelerated failure rate. This assembly had to be completely rebuilt in 2011, and no failure occurred since the new design has been implemented.


\subsection{{\sc Canopus}} 
\label{sub:canopusAIT}

\subsubsection{Optics} 
\label{ssub:optics}

One of the major difficulties in the original optical alignment of {\sc Canopus} was to adjust the focus of each optical path (LGS, NGS, Science). The constraints are: (a) the LGSWFS zoom should be able to span a range from approximately 87 to 140$\:$km (that is, covering all possible range to the sodium layer from zenith to an elevation of 40$\:$degrees); (b) the science focus is fixed by the GSAOI detector, which is not adjustable in focus; (c) the NGSWFS focus is fixed by the position of the mechanical assembly, which can be manually adjusted by few millimetres (see Fig. \ref{fig:gems_diagram}). Fine adjustments of the OAP position were also necessary to adjust the three focus simultaneously, while keeping the 2 arcmin field clear of any vignetting and the non-common path aberrations within the required level.

The LGS path throughput (from the entrance shutter of {\sc Canopus} to the LGSWFS CCD, not including the Quantum Efficiency of 0.8) was measured in the lab to be 35\%  at 589nm. The split is about 65\% for the LGSWFS itself (20 optical surfaces at 98\% each) and 55\% for {\sc Canopus} common path + WFS path. Admitedly, this is on the low side and could be improved in the future. The optical quality, including elements from the input focal plane calibration sources to the LGS WFS lenslets, is on the order of 250$\:$nm of astigmatism (averaged over the five LGS paths). Differential focus and astigmatism between the five paths were an important issue during the AIT and later on during commissioning. Differential focus was compensated by adjusting the individual collimators in the LGSWFS in 2012. Differential astigmatism cannot be corrected, and should be included in the NCPA procedure. 

The optical quality in the NGSWFS was also estimated: NGS spot sizes of about 0.3-0.4 arcsec in all three probes were measured when using diffraction-limited calibration sources. NGS spots are slightly elongated, most probably due to residual astigmatism on the order of 150$\:$nm rms, combined with defocus. However, there is no optical element in the NGSWFS path which can be used to compensate for residual aberrations. Due to design errors and alignment issues, the NGSWFS suffers from more than 2 magnitudes of sensitivity loss. Most of the light loss is happening at the injection of the light into the fiber, and the coupling between the fiber and the APDs. New APD modules were purchased and installed in 2011, providing a better fiber/APD coupling which resulted in a gain of about 1.5$\times$ in flux. A new fiber injection module has also been designed and implemented for one of the probes (C1) but failed to bring the expected improvement and was subsequently removed.

Finally, the image quality in the science path was measured with a high-order WFS in the lab, and then directly on the science camera when at the telescope (see Section~\ref{ssub:opra}). A fine adjustment of the output OAP was performed in order to reduce astigmatism in the science path. 
Without any NCPA compensation, the raw optical quality of the science path gives H band Strehl ratios on the order of 15\% to 30\% over the 2$\:$arcmin field. With the NCPA, this number goes up to about 90\%, as will be seen in paper II. 

The system end-to-end throughput (from outside the atmosphere to the GSAOI detector included) was measured to be 36\% in H (more details in \cite{carrasco2012results}), 21\% in J and 31\% in K, better than the initial design value of 23\% for all wavelengths.

\subsubsection{Non-common path aberration compensation} 
\label{ssub:opra}

Of the multitude of calibrations that have to be done with an AO system (and even more so with an MCAO system), the calibration and compensation of static non-common path aberrations (NCPA) is one of the most important. As the name says, those aberrations arise in the paths that are non-common, i.e. generally speaking after the light split between science and WFS paths. Science path aberrations are not seen by the AO WFS and thus not compensated; the WFS path aberrations are seen of course, therefore compensated, while they shouldn't be as they do not affect the science image directly. These aberrations are compensated by using WFS slope offsets. The difficulty consists in calibrating these aberrations: a wave-front sensing device in the science path is needed. The aberrations measured in the science path are compensated by adding --in software, e.g. using slope offsets-- the inverse aberrations to the AO WFS.

In GeMS, the problem is more complex; the goal is to compensate for NCPA over the entire field of view {\em simultaneously}. Because in the general case aberrations are not constant over the field of view, they also have to be calibrated and compensated depending on the position in the field of view. Several different methods were tried to perform this task; eventually, we settled on an improved version of the method proposed by \cite{kolb2006eliminating}. 
This novel approach \citep{gratadour2011tomographic}, called Tomographic Phase Diversity, is similar to the Phase Diversity + Tomography proposed by Kolb, except that instead of solving for the phase in each individual direction and then solving the tomography with the individual direction phases (to find the tomographic phase correction to apply to individual DMs), one solves directly in the volumetric phase space, using the many individual PSFs as input to the phase diversity process. This method provides better stability and has improved SNR properties compared to the original method proposed by Kolb. Results are given in paper II.


\subsubsection{Cooling}
\label{ssub:cooling}
A major engineering effort was required to re-design the thermal enclosures of the {\sc Canopus} electronics, particularly to manage the heat load of the Deformable Mirror Electronics (DME), 2900W accounting for about 70\% of the total 4100 W heat waste to be extracted from the instrument. Because the DME components are particularly sensitive to over temperatures this called for a complete and thorough redesign using new heat exchangers, high performance DC fans, compressed dry air, active valves and new telemetry to monitor the enclosure environment, electronics temperatures and any risk of condensation. The local turbulence in the bench is on the order of r$_0(500{\rm nm})=4\:$m, which proves that the thermal insulation is effective. 

\subsubsection{Mechanics} 
\label{ssub:mechanics}
Overall, the mechanisms and motors in the AO bench are performing well. The positioning reliability of the LGS WFS stepper motors was checked by taking measurements of DM0 to lenslet pupil registration when moving the bench between 0 and 54 degrees over 50 cycles. The motors performed reliably and under specification in those tests, keeping the average mis-registration below 4\% (peak-to-valley) of a subaperture in all beams. Residual flexure is compensated by a look-up table. No other flexure ---including differential flexures between the different paths --- was detected. Drift is mainly caused by temperature: the LGSWFS optical axis moves by approximately 200$\:$milliarcsec per 1$^o$C of temperature change. When working with the bench calibration sources, this precluded the utilisation of the TTM for centering the LGS WFS when operating with above a certain range of temperatures ($\Delta$T $\simeq$ 5 degrees Celsius), given that the TTM full range range is only 2.8 arcsec. A mechanical stage to adjust the position of the calibration sources along the drifting axis was added in order to compensate this issue.

During the design phase, special care was taken to reduce the impact of vibrations, using rigid, fixed optical mounts for instance. The level of vibrations measured in the lab, and when the final cooling solution was operational (see \ref{ssub:cooling}), was fully acceptable, at the level of 2 and 5$\:$milliarcsec rms along the two WFS axis. Vibrations measured on the telescope are slightly larger, on the order of 2 and 7$\:$milliarcsec rms, with some clearly identified peaks at 12 and 55$\:$Hz \citep{neichel2011kalman}. Occasionally, this goes up to 10$\:$milliarcsec rms, e.g. when the cryocooler are pumping hard to cool down an instrument.


\subsubsection{RTC and loops} 
\label{ssub:rtc}
The High-Order and TT loop behaviour (latency and bandwidth) were calibrated during the AIT period. Measuring the error transfer function on real signal (e.g. noise) is a very powerful tool to characterise the end-to-end properties of such dynamical systems made of optical, mechanical and electronic components. It allowed us to discover (and subsequently fix) a bug in which TT measurements were buffered and used with a one frame delay. Once this problem was fixed, an excellent agreement was found between model and experimental data.

The High-Order loop latency (defined here as the delay between the last pixel received from the LGS WFS CCDs and the last command applied to the DM) was measured to be 50$\:\mu s$. When adding 1.25$\:$ms of read-out time, this results in a total delay of 1.3$\:$ms. The DM response time was found to be negligible (except for failing actuators, see Section~\ref{ssub:dm0_actuator_failing}), which is what was expected from the manufacturer data. For the TT loop, the main factor limiting dynamical performance is the TT mirror mechanical behaviour. From manufacturer data (Physik Instrumente), the TTM has a -3dB point at 300Hz, which is in full agreement with our measurements. Overall, when running with maximum gains, the 0dB bandwidth (0dB point in the error transfer function) was measured to be approximately 53$\:$Hz for the high order loop and 40$\:$Hz for the TT loop.



\begin{figure*}
  \caption[GeMS first light: NGC288] {GeMS first light: NGC288 in H band}
  \begin{center}
  \includegraphics[width = 1.0\linewidth]{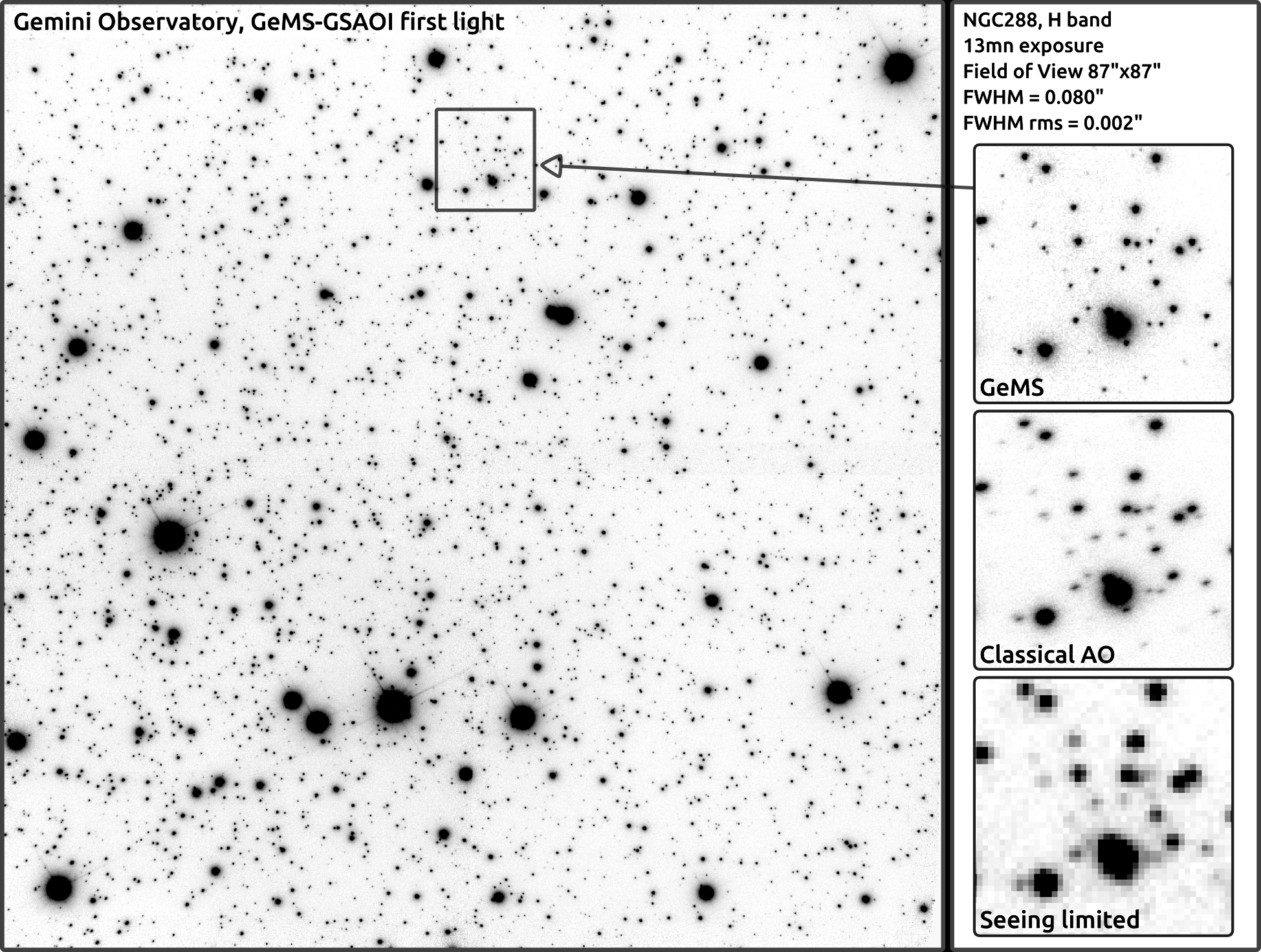}
  \end{center}
  \label{fig:ngc288}
\end{figure*}

\subsection{Issues and lessons learned} 
\label{sub:issues}

Not surprisingly, because GeMS was the first instrument of its kind, its development encountered a few issues, discussed below.

\subsubsection{WFS CCD pixel Modulation Transfer Function} 
\label{ssub:wfs_ccd_pixel_modulation_transfer_function}

As explained in section~\ref{sub:design_choices_and_trade_off}, there are only 2$\times$2 pixels per LGS WFS subaperture, each $1.38''$ in extent. The pixel extent was chosen as a compromise between spot clipping by the subaperture field of view (in case of bad LGS spot quality or bad seeing) and degradation of the spot Full-Width at Half Maximum (and thus the SNR) due to broadening of the spot by the detector pixel Modulation Transfer Function (MTF). Indeed, CCD pixels do not have abrupt edges. Through a phenomenon called pixel edge diffusion, there is a finite probability that a photon falling within the boundaries of a pixel be detected by a neighbouring pixel instead (see e.g. \cite{widenhorn2010charge}). 

By measuring the subaperture centroid gains (proportional to the FWHM, with a factor of proportionality depending on the assumed spot shape), and knowing the calibration source angular size, one can calculate the difference. Typical centroid gains obtained in the lab are 0.7, translating into equivalent FWHMs of $1.3''$ (this is subaperture dependent, but turned out to be relatively uniform). Given that the LGS calibration source size is $0.8''$, the degradation kernel is $1.0''$ which, after having eliminated other possible sources (e.g. defocus), we attributed to a MTF degradation by the detector. This value of $1''$ for the FWHM equivalent of the MTF degradation kernel, or about 2/3 of a pixel, is not uncommon and matches values measured in dedicated experiments with similar thinned detectors \citep{widenhorn2010charge,vandam2004performance}. This issue could not be remedied with the current EEV-CCD39. An obvious solution would be to upgrade the detectors to low readout noise larger arrays, thus smaller pixels to sample properly the LGS spots. Such an upgrade is not considered to date, primarily because there are more serious issues to correct first.

\subsubsection{BTO design, LGS spot optical quality and brightness} 
\label{ssub:lgs_spot_optical_quality_and_brightness}
Although the BTO is made of relatively simple optics such as planar mirrors, lenses, beam splitters, and polarisation optics, it has proven to be a fairly complex system to align and optimize. One of the main issue that was encountered with the BTO was the location of the fast tip/tilt mirrors (FSM) or arrays (FSA) used to compensate for the fast up-link seeing. In GeMS, these five TT platforms are not located in a pupil plane (the LLT primary mirror), inducing a continuous jitter of the laser beam footprints on the LLT primary mirror. When static alignment of the five beams on the LLT is not perfectly done, i.e. when the five beams are not perfectly superimposed on the LLT, the risk of vignetting one or more of the beams is high. Moreover, some variations of the spot quality between the beams is observed, on the order of 0.1 to 0.2 arcseconds. This effect is attributed to LLT pupil aberrations and mainly caused by the LLT OAP mounting issues. Finally, not only will projected laser power and spot size per LGS vary over time, but the Rayleigh beam footprints on each LGS WFS (called fratricide) will also change rapidly, making it impossible to ever subtract the Rayleigh background from the LGS WFS frames as well as creating all sorts of spurious effects for the AO reconstructor.

\subsubsection{Failure of actuators on DM0} 
\label{ssub:dm0_actuator_failing}
When DM0 was first installed in {\sc Canopus}, all its actuators were functional. Over two years of AIT work in the lab (2008/2009), three actuators failed --- i.e. either they did not react or reacted very slowly. This failure mode is a feature of the DM itself and not of its power supplies. After moving {\sc Canopus} to Cerro Pach{\'o}n, actuators started failing more rapidly: Six months later, 16 more actuators were non-functional, and an actuator was lost every ten days in average. Although DM4.5 and DM9 did have some dead actuators, they did not show such an accelerated degradation as DM0. Entering the GeMS shutdown during the winter 2011, it was thus decided to replace DM0 with DM4.5, and to replace DM4.5 with a flat mirror. 

This has some side effects; positive ones were that it would make the control easier (two instead of three DMs) and that the static shape of DM4.5 was better than DM0, which showed some cylinder due to ageing. Negative ones were to reduce somewhat the expected system performance, given that the total number of active actuators was reduced from 684 to 360, and that the compensation of altitude layer was now effectively handled solely by DM9, with a rather modest actuator pitch of 1$\:$m. 


\begin{figure}
  \caption[NGC6369 with GMOS] {GeMS and GMOS: NGC6369 at R band}
  \begin{center}
  \includegraphics[width = 1.0\linewidth]{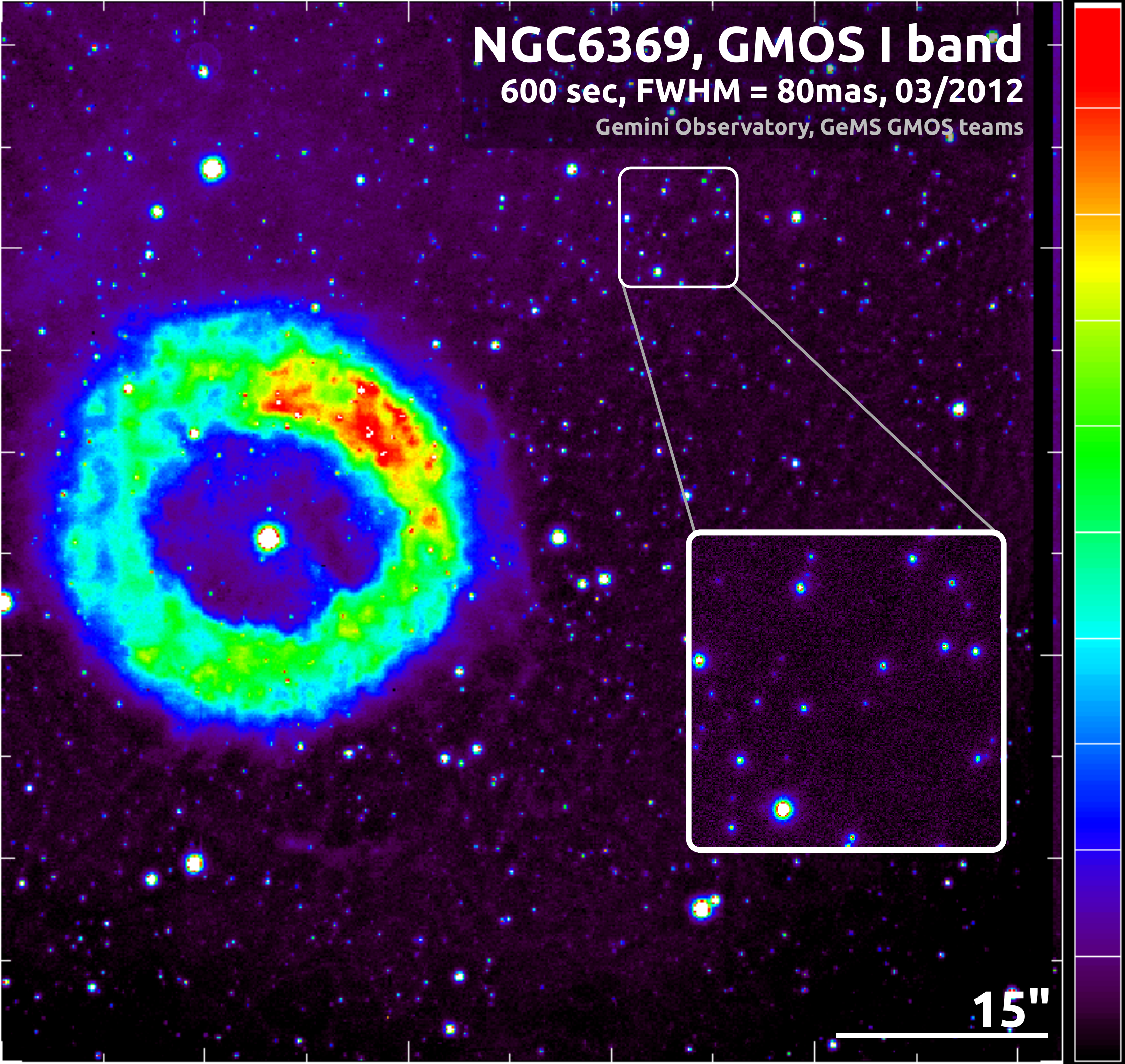}
  \end{center}
  \label{fig:ngc6369}
\end{figure}

\subsubsection{NGS WFS APDs feed} 
\label{ssub:ngs_wfs_apds_feed}
APD-based Tip-Tilt quad cell WFS are the norm in LGS AO systems. {\sc Strap}, an APD-based system developed by ESO \citep{bonaccini1998adaptive}, is in use at the ESO VLTs/VLTI, at the Keck I and II telescopes and at Gemini North amongst others. Because of the need to have three TT WFS with adjustable positions within a 2 arcmin FoV, {\sc Strap} was not an option for {\sc Canopus}. A three-probe system was designed by EOST, using focal plane pyramids to dissect the focal plane image, then direct it to fiber-fed APDs. These systems proved extremely difficult to align:  they had to be very compact to fit and avoid collisions in the NGS WFS focal plane, which thus prevented implementing the necessary alignment adjustments. Significant effort has been applied to upgrade these systems, with little success. A total redesign based on a single large focal plane array is being planned (see paper II), which should allow GeMS to reach a NGS limiting magnitude of R=18.5, as was originally specified.


\subsubsection{Differential field distortions Science/NGS} 
\label{ssub:differential_field_distortions_science_ngs}
Two-off-axis parabola systems are widely used in AO. They provide clean pupil re-imaging, with little pupil distortion. They transport the focal plane with very little aberrations over the generally small AO field of view. However they introduce a significant amount of distortion in the output focal plane. In the science focal plane, field distortions have minor consequences, since they can be calibrated out. In the TT NGS focal plane, this has serious consequences, the most severe of which is that the star constellation will deform when dithering (dithering is the normal mode of observation in the infrared). The TT WFS had been designed with probes \#1 and \#2 mounted on top of probe \#3. The intent was to be able to dither with a motion of probe \#3 only, and thus not deform the constellation, to be able to stack science images without having to correct for plate scale between them. The field distortion prevents that. In fact the distortion is so large that dither of 10 arcsec or so will induce differential motion on the order of 0.1-$0.2''$ between probes, which makes operation impossible and mandate going through another acquisition to centre the probes on their respective stars.

The proposed redesign of the TT WFS mentioned above (see also paper II) will solve this problem, as the distortion model can be easily incorporated into the positioning model for each guide star on the focal plane array.


\subsubsection{Lessons learned} 
\label{ssub:lessons_learned}

In summary, to be beneficial to future instrument designers, here is a top level view of what worked and what didn't and, where we to build GeMS again today, what we would do the same and what we wouldn't.

Below are the items that caused the most trouble, either because they are limiting performance and/or because they caused very large overheads during AIT and/or commissioning (note that at the time of the GeMS design there was no alternative for most of these choices). All of these issues have been discussed at length earlier in this paper.

\begin{itemize}
  \item LGSWFS Quadcells, for two reasons: Pixel MTF (performance degradation) and centroid gains (performance degradation and huge calibration burden). Today's alternative is to use large EMCCDs to adequately sample Shack-Hartmann spots.
  \item NGSWFS probes: Beware of fiber feeds, mechanism (re)positionnings and guide star catalogue coordinate errors. Today's alternative is to use detector array(s): No moving parts, less optical elements should result in better performance, much simpler calibrations and a huge simplification of acquisition. If this is done properly, one can probably live with the distortion introduced by the two off-axis parabola relay.
  \item Laser center launch, for two reasons: First, the fratricide turned out to be a real problem. In GeMS, but probably more generally, the Rayleigh scattering can not be calibrated out. Second, because it implies a more complicated BTO relay, with many more optical elements and motors. In fact, the whole BTO, because of its complexity, has implied huge calibration overheads (e.g. constellation alignment). The lesson here is to simplify the BTO design as much as possible. Today's alternative is to use more compact lasers and/or side launch.
  \item Laser: Even though it is a technological feat, GeMS's laser is a very large, costly and complex system. Today's alternative are Raman fiber lasers or Optically Pumped Semiconductor Lasers.
  \item Higher conjugation altitude of high DM.
\end{itemize}

What worked well, and would be done the same way:
\begin{itemize}
  \item Optical design and instrument packaging,
  \item Field of view and constellation geometry,
  \item Reduced actuator density on high altitude DM,
  \item AIT in house and commissioning format (see paper II),
  \item Strong in-house AO team that take control of high level software and control.
\end{itemize}

Lessons learned have also been discussed at length in \cite[Sect.3]{rigaut2011gems}.




\begin{figure*}
  \caption[The antennae galaxies with GeMS and the HST] {Gems and HST complementarity: Different wavelengths, different views, same resolution}
  \begin{center}
  \includegraphics[width = 1.0\linewidth]{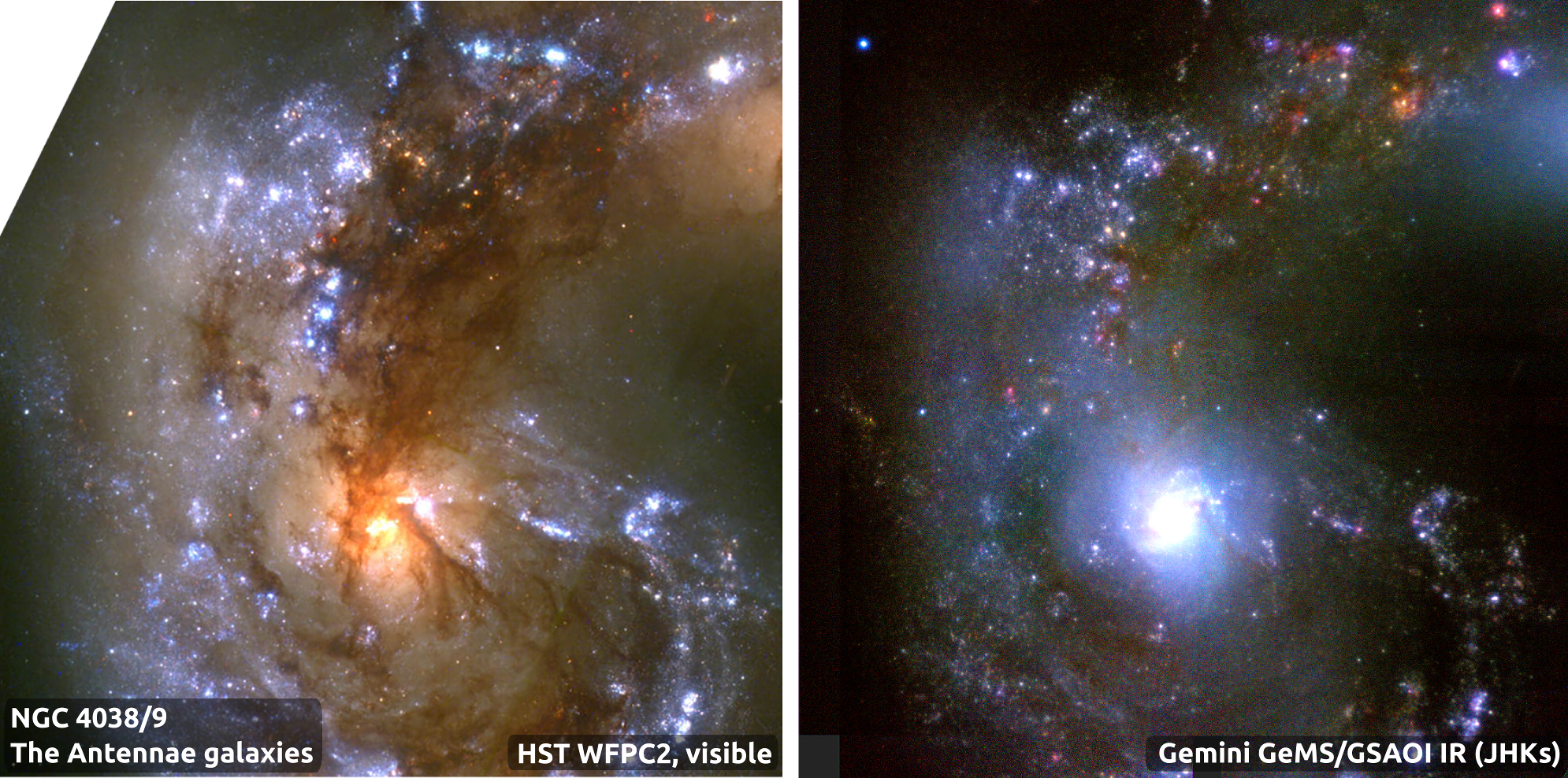}
  \end{center}
  \label{fig:antennae}
\end{figure*}

\section{First light} 
\label{sec:first_light}

Paper II describes in detail the commissioning, operation and performance of GeMS. This section only provides a summary of first light results. 

Commissioning took place over the course of 2011 and 2012. The so-called ``first light'' image was obtained on December 16 2011 on the globular cluster NGC288, and is shown in figure~\ref{fig:ngc288}. The seeing was $0.7''$ on this night, close to the median seeing for the site. This image is taken at 1.65 microns (H band) and has a field-of-view of $87''\times87''$. It is a combination of 13 images of 60 seconds each. The average full-width at half-maximum is slightly below 80 milliarcsec, with a variation of 2 milliarcsec across the entire image field of view. Insets on the right show a detail of the image (top), an image of the same region with classical AO (middle; this has been generated from the top MCAO-corrected image and assumes using the star at the upper right corner as the guide star), and seeing-limited observations (bottom). The pixel size in the seeing-limited image was chosen to optimize signal-to-noise ratio while not degrading angular resolution\footnote{Keeping the same pixel size as the MCAO image would have resulted in a lot of noise in the seeing limited image, hence to present a fair comparison, we choose to use larger pixels, also more realistic, to generate the seeing limited image.}. North is up, East is right. Strehl ratios across the image are on the order of 15 to 20\%. The relatively large FWHM, compared with what has been obtained more recently, can be explained in part because the plate scale and the focus stabilisation loops were not closed. Nevertheless, the nicely packed PSF, approximately Lorentzian in shape and without marked halo, is extremely uniform across the 87 arcsec field, demonstrating the very point of MCAO. 

Figure~\ref{fig:ngc6369} shows a single 600 second exposure of the planetary nebula NGC6369 acquired with the Gemini Multi-Object Spectrograph (GMOS) in the red, at I band (about 830nm in this case) on March 14, 2012. The field of view is $75''\times75''$. The natural seeing at the time was between 0.3 and $0.4''$; the FWHM of the corrected image over the displayed field is between 80 and 90 milliarcsec. Using GMOS with {\sc Canopus} has never been very high in the observatory priorities as it was believed performance was going to be marginal. This image, with about 80$\:$milliarcsec FWHM over most of the 2 arcmin field of view unvignetted by {\sc Canopus}, proves that when the seeing cooperates, GeMS can deliver down the red part of the visible spectrum. This image was the best obtained with GMOS though. Under median seeing conditions, GeMS provided typically a factor of 2 to 2.5 improvement in FWHM, which is roughly what was expected (i.e. slightly better than GLAO).

\cite{rigaut2012gems,rigaut2011gems} report on additional first light results: FWHM and Strehl ratio uniformity maps and a preliminary error budget; \cite{rigaut2012gems} goes on with an identification of the factors limiting performance at the time of the first light: Photon return (i.e. mostly due to low laser power projected on-sky) and generalised fitting (a consequence of the missing DM4.5). Static aberrations were also a problem at the time of first light but has been fixed since then. Finally, the same paper gives a preliminary analysis of the astrometric performance, showing that sub-milliarcsec accuracy can be readily achieved, and that MCAO is not introducing uncontrolled terms to the astrometric error budget. This was later confirmed by \cite{lu2013astrometry}, who finds that 0.1-0.2 milliarcsec astrometric accuracy can be reached.

Since first light, GeMS has acquired many new stunning images. A collection of legacy images taken on various objects ranging from the Orion Nebula to the galaxy cluster Abell 780, through globular \& open clusters has recently being published on the Gemini observatory website (\url{http://www.gemini.edu/node/12020}). Figure~\ref{fig:antennae} shows one of these legacy images: the antennae galaxies (NGC4038/39) as seen by the Hubble Space Telescope (left, composite visible image) and GeMS (right, composite infrared image). Because of the amount of dust, largely opaque to visible light, the view offered by HST and by GeMS are significantly different. GeMS's infrared view, at an angular resolution similar or slightly better than HST in the visible, provides extremely useful complimentary information to the study of astrophysical objects.

In 2013, \cite{davidge2013haffner} and \cite{zyuzin2013vela} published the first science papers using GeMS data.


\section{Conclusion} 
\label{sec:conclusion}

Over ten years of efforts and one year of commissioning culminated in December 2011 with the first GeMS/GSAOI science images. This paper is the first paper in a two-part GeMS review. We gave an overview of the history, design and trade-offs, provided a description of the system assembly, integration and tests, and commented on the issues \& lessons learned. Paper II reports on GeMS commissioning, performance and  operation on sky, and GeMS upgrade plans.

In conclusion, {\em GeMS is fulfilling the promise of wide-field AO}. Over the 85$\times$85 arcsec field of view of GSAOI, images with FWHM of 80$\:$milliarcsec, with exquisite uniformity, are typically obtained under median seeing conditions. Strehl ratio of 40\% in H band have been obtained, which we believe are the highest to date with a LGS-based AO system on a large telescope, which are typically limited by focus anisoplanatism at this wavelength.

Finally, GeMS, and the experience acquired from it, is also crucial for the design of the AO systems of the future generation of extremely large telescopes (GMT, TMT, E-ELT).

GeMS/GSAOI is a unique instrument, and will no doubt deliver first class science.

\section*{Acknowledgments}

GeMS was a large instrumentation project. In the course of the last 13 years, it involved people from many different disciplines. The authors would like to recognise the contribution and thank 
Claudio Arraya, 
Rodrigo Carrasco, 
Felipe Colazo, 
Fabian Collao, 
Paul Collins, 
Herman Diaz, 
Sarah Diggs, 
Matt Doolan, 
Anne Drouin, 
Michelle Edwards, 
Vincent Garrel, 
Fred Gillett, 
Alejandro Gutierrez, 
Mark Hunten, 
Stacy Kang, 
Matteo Lombini 
Ariel Lopez, 
Claudio Marchant, 
Eduardo Marin, 
Peter McGregor, 
Brian Miller, 
Cristian Moreno, 
Matt Mountain, 
Peter Pessev, 
Rolando Rogers, 
Jean-Ren{\'e} Roy, 
Andrew Serio, 
Doug Simons, 
Chad Trujillo, 
Cristian Urrutia, 
Jan van Harmelen, 
Vicente Vergara, 
Tomislav Vucina, 
Peter J. Young, 
and the panel members of the many reviews GeMS went through.

Based on observations obtained at the Gemini Observatory, which is operated by the 
Association of Universities for Research in Astronomy, Inc., under a cooperative agreement with the NSF on behalf of the Gemini partnership: the National Science Foundation (United States), the National Research Council (Canada), CONICYT (Chile), the Australian Research Council (Australia), Minist\'{e}rio da Ci\^{e}ncia, Tecnologia e Inova\c{c}\~{a}o (Brazil) and Ministerio de Ciencia, Tecnolog\'{i}a e Innovaci\'{o}n Productiva (Argentina).


\bibliographystyle{mn2e}

\bibliography{ao,gems}

\label{lastpage}

\end{document}